\documentclass{ws-procs9x6}

\newcommand{\be}{\begin{equation}}
\newcommand{\ee}{\end{equation}}

\makeatletter
\newbox\slashbox \setbox\slashbox=\hbox{\large$/$}
\def\pslash#1{\setbox\@tempboxa=\hbox{$#1$}
  \@tempdima=0.5\wd\slashbox \advance\@tempdima 0.5\wd\@tempboxa
  \copy\slashbox \kern-\@tempdima \box\@tempboxa}

\def\@cite#1#2{{#1\if@tempswa , #2\fi}}
\def\@citex[#1]#2{%
\if@filesw\immediate\write\@auxout{\string\citation{#2}}\fi
\leavevmode\unskip$^{\,\scriptstyle\@cite{\@collapse{#2}}{#1}}$}
\def\CITE{\@ifnextchar[{\@tempswatrue\@CITEX}{\@tempswafalse\@CITEX[]}}
\let\onlinecite\CITE
\def\@CITEX[#1]#2{%
\if@filesw\immediate\write\@auxout{\string\citation{#2}}\fi
\leavevmode\unskip\ \@cite{\@collapse{#2}}{#1}}
\def\@collapse#1{{%
\let\@temp\relax
\@tempcntb\@MM
\def\@citea{}%
\@for \@citeb:=#1\do{%
\@ifundefined{b@\@citeb}%
{\@temp\@citea{\bf ?}%
\@tempcntb\@MM\let\@temp\relax
\@warning{Citation `\@citeb ' on page \thepage\space undefined}}%
{\@tempcnta\@tempcntb \advance\@tempcnta\@ne
\edef\MyTemp{\csname b@\@citeb\endcsname}%
\def\@tempa{\@temptokena=\bgroup}%
\afterassignment\@tempa %
\@tempcntb=0\MyTemp\relax}%
\ifnum\@tempcntb=0\relax%
\@tempcntb=\@MM
\@citea\MyTemp
\let\@temp = \relax
\else %
\edef\@tempd{\number\@tempcntb}%
\ifnum\@tempcnta=\@tempcntb %
\ifx\@temp\relax %
\edef\@temp{\@citea\@tempd}%
\else
\edef\@temp{\hbox{--}\@tempd}%
\fi
\else %
\@temp\@citea\@tempd
\let\@temp\relax
\fi
\fi
}%
\def\@citea{, }%
}%
\@temp %
}}%

\makeatother

\begin{document}
\title{Monopoles in Real Time for Classical U(1) Gauge Field Theory}
\author{Tam\'as S. Bir\'o}
\address{KFKI, Research Institute for Particle and Nuclear Physics, H-1525 Budapest, Hungary}
\author{Harald Markum and Rainer Pullirsch}
\address{Atominstitut, Technische Universit\"at Wien, A-1040 Vienna, Austria}
\maketitle

\abstracts{
        U(1) gauge fields are decomposed into a monopole and photon part
        across the phase transition from the confinement to the Coulomb
        phase.  We analyze the leading Lyapunov exponents of such
        gauge field configurations on the lattice
        which are initialized by quantum Monte Carlo simulations.
        We observe that the monopole field carries the same
        Lyapunov exponent as the original U(1) field.
        As a long awaited result, we show that monopoles are created and
        annihilated in pairs as a function of real time in excess to a fixed
        average monopole number.
}

\section {Monopole and photon part of U(1) theory}

We are interested in the relationship between monopoles
and classical chaos across the phase transition
at $\beta_c \approx 1.01$.
We begin with a $4d$ U(1) gauge theory described by the 
Euclidean action
$ S \lbrace U_l \rbrace = \beta \: \sum_p (1 - \cos \theta_p )$,
where $U_l = U_{x,\mu} = \exp (i\theta_{x,\mu}) $ and
$
  \theta_p =
 \theta_{x,\mu} +
 \theta_{x+\hat{\mu},\nu} -
 \theta_{x+\hat{\nu},\mu} -
 \theta_{x,\nu}\ . $
Following Ref.~\onlinecite{StWe92}, we have
factorized our gauge configurations
into monopole and photon fields.
The U(1) plaquette angles $\theta_{x,\mu\nu}$ are decomposed into the
``physical'' electromagnetic flux through the plaquette
$\bar \theta_{x,\mu\nu}$ and a number $m_{x,\mu\nu}$ of Dirac strings
through the plaquette
\begin{equation} \label{Dirac_string_def}
 \theta_{x,\mu\nu} = \bar \theta_{x,\mu\nu} + 2\pi\,m_{x,\mu\nu}\ ,
\end{equation}
where $\bar \theta_{x,\mu\nu}\in (-\pi,+\pi]$ and
$m_{x,\mu\nu} \ne 0$ is called a Dirac plaquette.

\section{Classical chaotic dynamics from quantum Monte Carlo
         initial states}

Chaotic dynamics in general is characterized by the
spectrum of Lyapunov exponents. These exponents, if they are positive,
reflect an exponential divergence of initially adjacent configurations.
In case of symmetries inherent in the Hamiltonian of the system
there are corresponding zero values of these exponents. Finally
negative exponents belong to irrelevant directions in the phase
space: perturbation components in these directions die out
exponentially. Pure random gauge fields on the lattice show a characteristic
Lyapunov spectrum consisting of one third of each kind of
exponents.\cite{BOOK}
Assuming this general structure of the Lyapunov spectrum we
present here its magnitude only, namely the maximal
value of the Lyapunov exponent, $L_{{\rm max}}$.

The general definition of the Lyapunov exponent is based on a
distance measure $d(t)$ in phase space,
\begin{equation}
L := \lim_{t\rightarrow\infty} \lim_{d(0)\rightarrow 0}
\frac{1}{t} \ln \frac{d(t)}{d(0)}.
\end{equation}
In case of conservative dynamics the sum of all Lyapunov exponents
is zero according to Liouville's theorem, $\sum L_i = 0$.
We utilize the gauge invariant distance measure consisting of
the local differences of energy densities between two $3d$ field configurations
on the lattice:
\begin{equation}
d : = \frac{1}{N_P} \sum_P\nolimits \, \left| {\rm tr} U_P - {\rm tr} U'_P \right|.
\end{equation}
Here the symbol $\sum_P$ stands for the sum over all $N_P$ plaquettes,
so this distance is bound in the interval $(0,2N)$ for the group
SU(N). $U_P$ and $U'_P$ are the familiar plaquette variables, constructed from
the basic link variables $U_{x,i}$,
\begin{equation}
U_{x,i} = \exp \left( aA_{x,i}T \right)\: ,
\end{equation}
located on lattice links pointing from the position $x=(x_1,x_2,x_3)$ to
$x+ae_i$. The generator of the group U(1) is
$T = -ig$ and $A_{x,i}$ is the vector potential.
The elementary plaquette variable is constructed for a plaquette with a
corner at $x$ and lying in the $ij$-plane as
$U_{x,ij} = U_{x,i} U_{x+i,j} U^{\dag}_{x+j,i} U^{\dag}_{x,j}$.
It is related to the magnetic field strength $B_{x,k}$:
\begin{equation}
U_{x,ij} = \exp \left( \varepsilon_{ijk} a^2 B_{x,k} T \right).
\end{equation}
The electric field strength $E_{x,i}$ is related to the canonically conjugate
momentum $P_{x,i} = a \dot{U}_{x,i}$ via
\begin{equation}
E_{x,i} = \frac{1}{g^2a^2} \left( T \dot{U}_{x,i} U^{\dag}_{x,i} \right).
\end{equation}

The Hamiltonian of the lattice gauge field system can be casted into
the form
\begin{equation}
H = \sum \left[ \frac{1}{2} \langle P, P \rangle \, + \,
 1 - \frac{1}{4} \langle U, V \rangle \right].
\end{equation}
Here the scalar product stands for
$\langle A, B \rangle = \frac{1}{2} {\rm tr} (A B^{\dag} )$.
The staple variable $V$ is a sum of triple products of elementary
link variables closing a plaquette with the chosen link $U$.
This way the Hamiltonian is formally written as a sum over link
contributions and $V$ plays the role of the classical force
acting on the link variable $U$. 

We prepare the initial field configurations
from a standard four dimensional Euclidean Monte Carlo program on
a $12^3\times 4$ lattice varying the inverse gauge coupling $\beta$.\cite{SU2}
We relate such four dimensional Euclidean
lattice field configurations to Minkow\-skian momenta and fields
for the three dimensional Hamiltonian simulation
by selecting a fixed time slice of the four dimensional lattice.

\section{Chaos and confinement }

\begin{figure*}[ht]
\centerline{{\hspace*{5mm}\psfig{figure=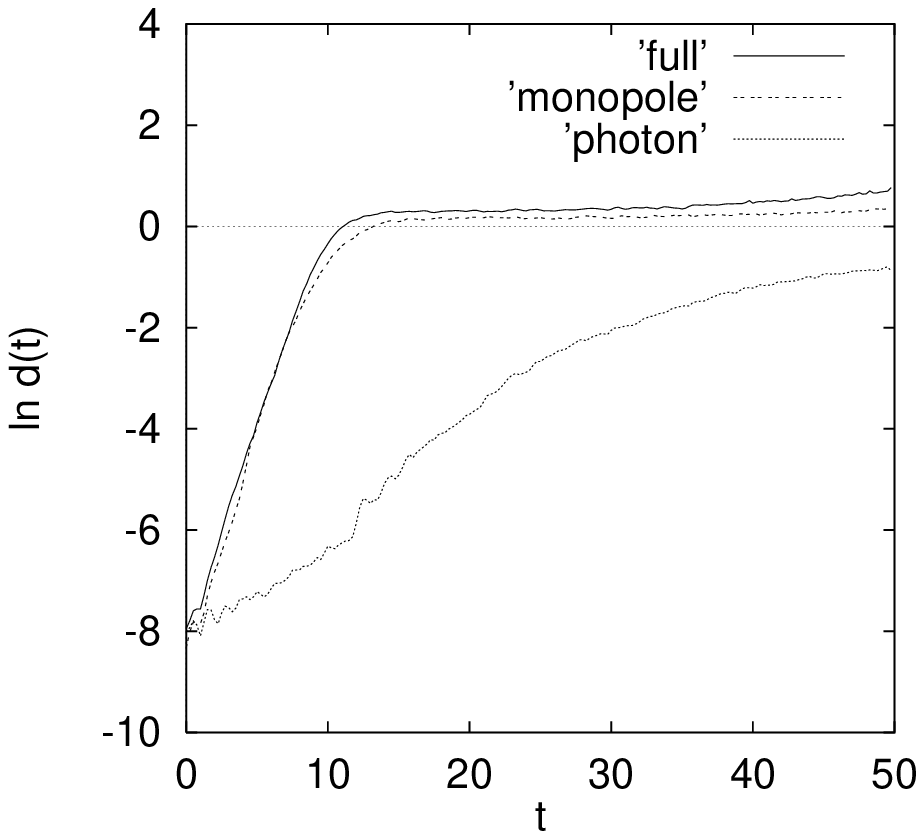,width=5.5cm}}\hspace{5mm}
{\psfig{figure=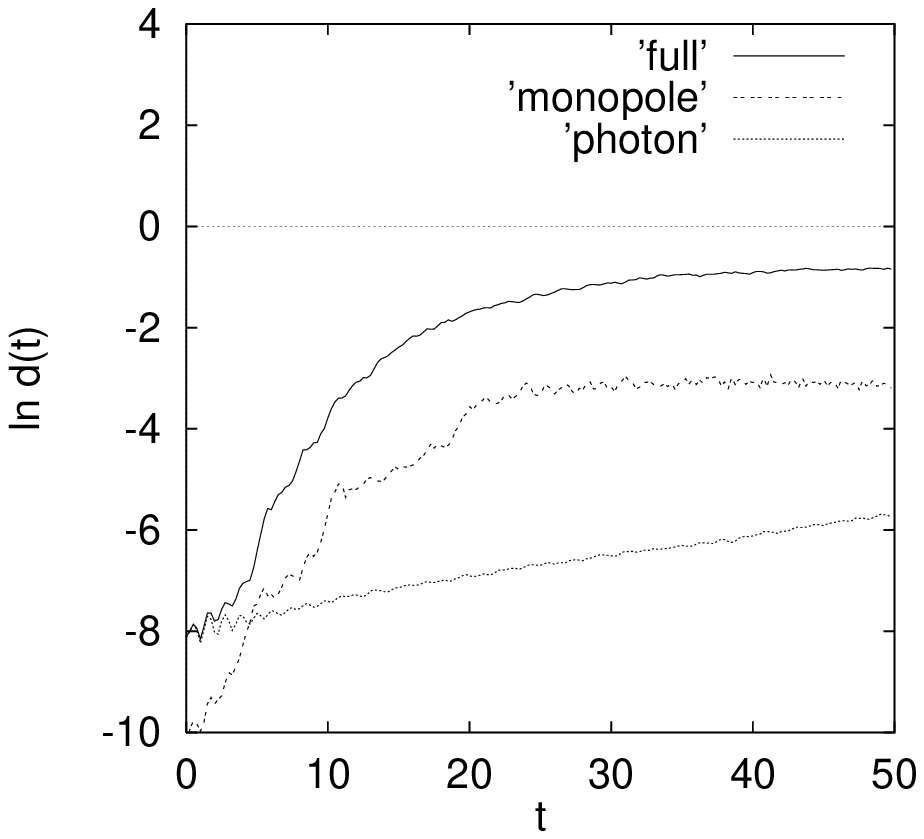,width=5.5cm}}}
\vspace*{-0mm}
\caption{
  Exponentially diverging distance in real time of initially adjacent U(1) field
  configurations on a $12^3$ lattice prepared at $\beta=0.9$ in the
  confinement (left) and at $\beta=1.1$ in the Coulomb
  phase (right).
\vspace*{-0mm}
\label{Fig2}
 }
\end{figure*}

\noindent We start the presentation of our results with a characteristic example
of the time evolution of the distance between initially adjacent
configurations. An initial state prepared by a standard four dimensional
Monte Carlo simulation is evolved according to the classical Hamiltonian dynamics
in real time. Afterwards this initial state is rotated locally by
group elements which are chosen randomly near to the unity.
The time evolution of this slightly rotated configuration is then
pursued and finally the distance between these two evolutions
is calculated at the corresponding times.
A typical exponential rise of this distance followed by a saturation
can be inspected in Fig.~\ref{Fig2} from an example of U(1) gauge theory
in the confinement phase and in the Coulomb phase.
While the saturation is an artifact of
the compact distance measure of the lattice, the exponential rise
(the linear rise of the logarithm)
can be used for the determination of the leading Lyapunov exponent.
The left plot exhibits that in the confinement phase the original
field and its monopole part have similar Lyapunov exponents whereas
the photon part has a smaller $L_{max}$. The right plot in the Coulomb
phase suggests that all slopes and consequently the Lyapunov
exponents of all fields decrease substantially.

\section{Real-time evolution of monopoles}

\begin{figure*}[htb]
\includegraphics[width=45mm,height=54mm,angle=-90]{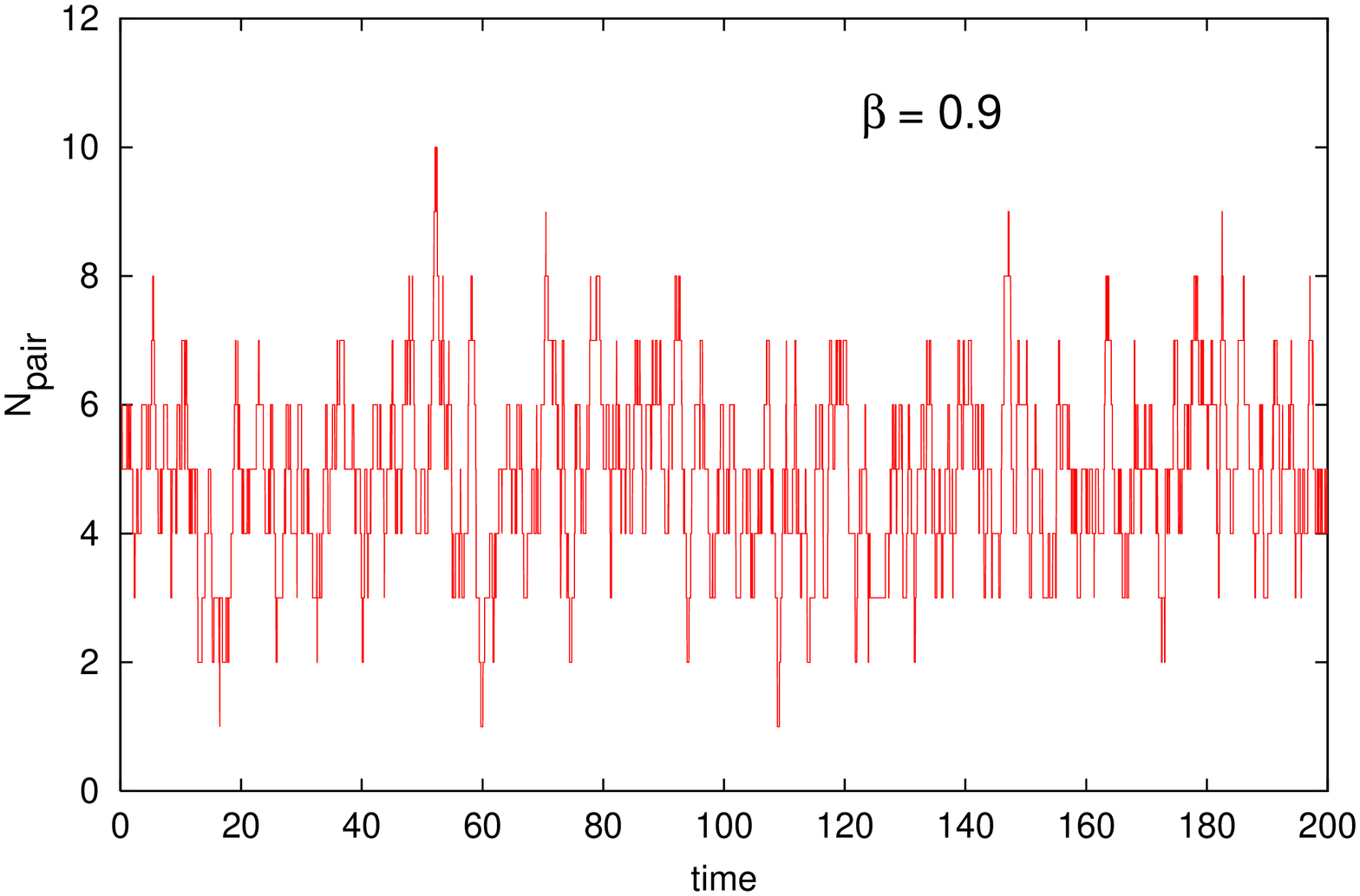}\hspace{5mm}
\includegraphics[width=45mm,height=54mm,angle=-90]{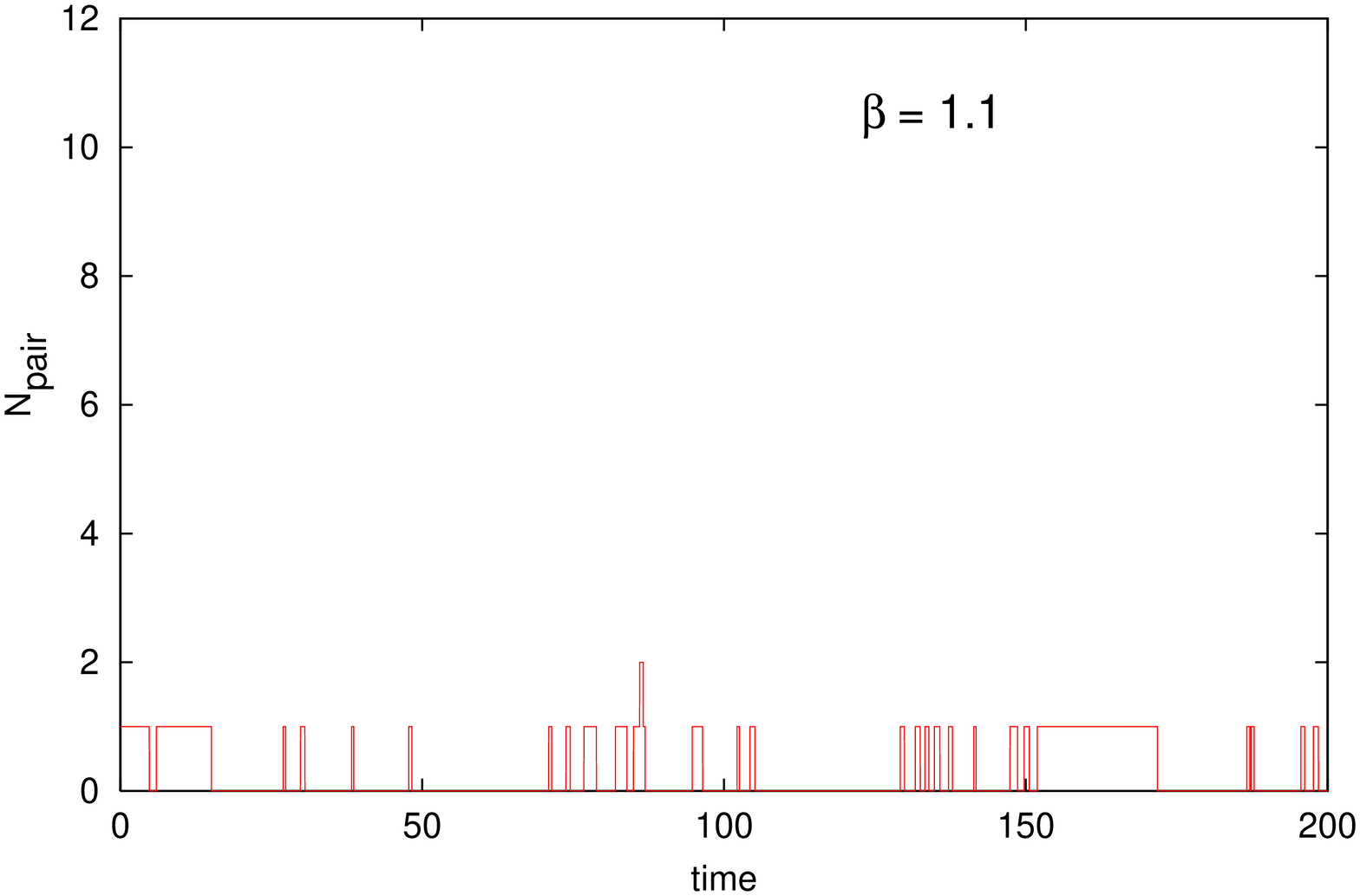}
\caption{The number of monopole pairs as a function of real time evolution
at $\beta=0.9$
(confined, left) and at  $\beta=1.1$ (Coulomb phase, right).}
\label{pair09_11}
\end{figure*}
\begin{figure*}[ht]
\includegraphics[width=45mm,height=54mm,angle=-90]{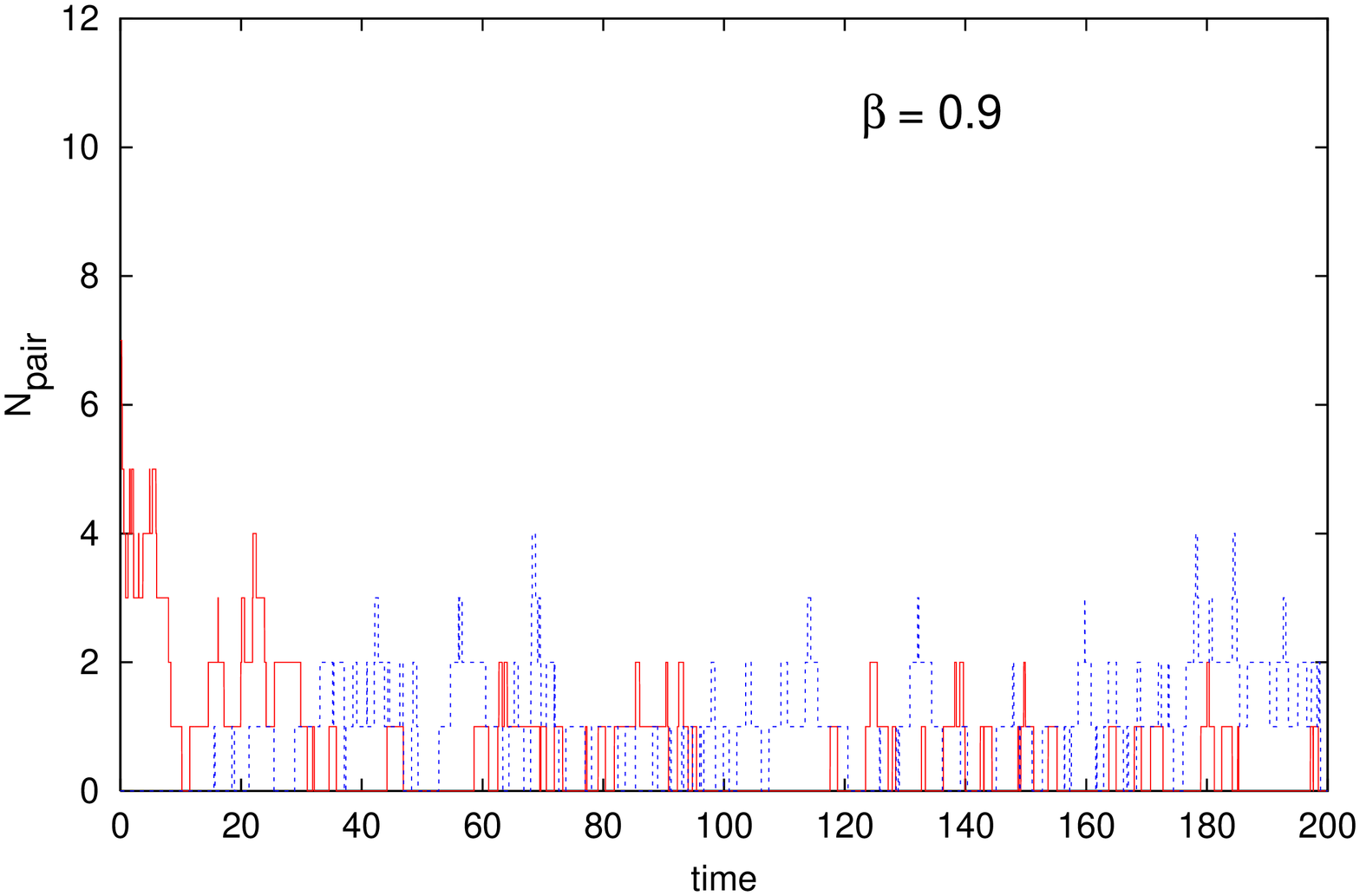}\hspace{5mm}
\includegraphics[width=45mm,height=54mm,angle=-90]{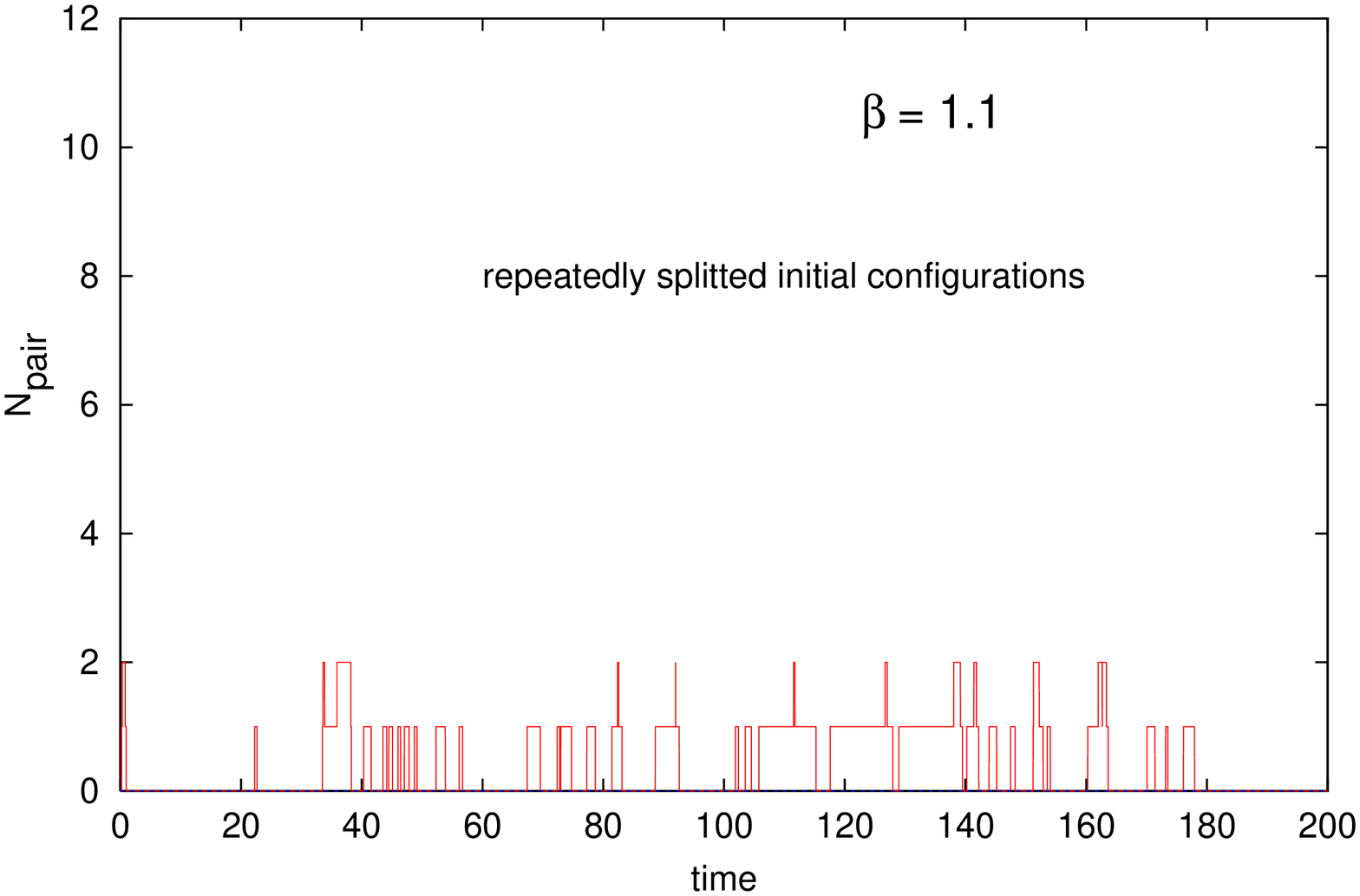}
\caption{Monopole pairs in the decomposed field at $\beta=0.9$ (left)
and at $\beta=1.1$ (right). In the latter case the decomposition was
repeated until complete absence of monopoles in the photon part.
Full lines correspond to
the monopole, dashed lines to the photon (residual) configuration.}
\label{asplit09_11}
\end{figure*}

\noindent An important  subject of the present study is the evolution of monopoles in
real time.\cite{U1} The outcome is displayed in Fig.~\ref{pair09_11} and shows that
their number stays constant on average with creation and annihilation of
a few monopole pairs only. The fluctuations decrease with increasing
inverse gauge coupling $\beta$. A decrease of the total number of monopoles
takes place towards the Coulomb phase.

The real-time evolution of the monopole content in the decomposed U(1)
field is depicted in Fig.~\ref{asplit09_11}. Although there are less
monopoles in the photon field than in the monopole field, their
fluctuations are comparable in the confining phase. On the right plot,
made for the Coulomb phase at $\beta=1.1$, additionally the decomposition
to photon and monopole fields has been repeatedly applied until the
photon part had no monopole in it. This zero value is conserved
during the real-time evolution.

\end{document}